\documentclass[12pt,aps,floats,showpacs,amssymb,tightenlines]{revtex4}

\usepackage{color}
\usepackage{graphicx}
\usepackage{amsmath}
\usepackage{amsfonts}
\usepackage{amscd}
\usepackage{epsfig}
\usepackage{amssymb}
\usepackage{tabularx}

\begin{document}

\title{On the dynamics of thin shells of counter rotating
particles}
\author{Reinaldo J. Gleiser} \email{gleiser@fis.uncor.edu} \author{Marcos A. Ramirez}
\affiliation{Facultad de Matem\'atica,
Astronom\'{\i}a y F\'{\i}sica, Universidad Nacional de C\'ordoba,
Ciudad Universitaria, (5000) C\'ordoba, Argentina}
\email{gleiser@fis.uncor.edu}

\begin{abstract}
In this paper we study the dynamics of self gravitating spherically
symmetric thin shells of counter rotating particles. We consider all possible velocity distributions for the particles, and show that the equations of motion by themselves do not constrain this distribution. We therefore consider the dynamical stability of the resulting configurations under several possible processes. This include the stability of static configurations as a whole, where we find a lower bound for the compactness of the shells. We analyze also the stability of the single particle
orbits, and find conditions for ``single particle evaporation''. Finally, in the case of a shell with particles whose angular momenta are restricted to two values, we consider the conditions for stability under splitting into two separate shells. This analysis leads to the conclusion that under certain
conditions, that are given explicitly, an evolving shell may split
into one or more separate shells. We provide explicit examples to illustrate this phenomenon. We also include a derivation of the thick to thin shell limit for an Einstein shell that shows that the limiting distribution of angular momenta is unique, covering continuously a finite range of values.
\end{abstract}

\maketitle

\section{Introduction}

Constructing explicit solutions of Einstein's equations with matter that satisfies a physically acceptable equation of state, or, even acceptable energy conditions, has proved to be in general a quite difficult task. The problem may be simplified by the introduction of symmetries, for instance assuming spherical symmetry, and the imposition of extreme equations of state, such as, e.g., constant energy density. A further important geometrical simplification concerns the assumption that the support of the region where the matter density is non vanishing is an embedded hypersurface in the space time manifold, that is, that the matter is restricted to a {\em shell}. In some applications, such as the recent brane models, this is an essential assumption of the theory. In other cases, such as those where astrophysical applications are in mind, they should be considered as simplifying limits, and therefore, it is quite important to establish what are the configurations for which the shell is an idealized limit, and whether the resulting configurations are stable. As regards the simplifications on the type of matter, a particularly interesting class, because of its relevance in astrophysics, is the assumption that it is made out of collisionless particles, where each particle follows a geodesic of the mean field generated, at least in part, by the rest of the particles, and where the evolution of the gravitational and matter fields is governed by the Einstein - Vlasov equations \cite{Andreas}. The problem here is again that the general equations turn out to be complicated, and one must resort to further simplifying assumptions in order to get definite answers. This has led to the consideration of spherically symmetric models, where the particle trajectories are also restricted. Two of these restrictions have proved to be particularly useful. The first is the restriction to radial motions in the {\em dust} models. The other corresponds to the assumption that at each point there is a radially comoving frame, where the radial velocities of all particles vanish. In the case of spherical symmetry, this implies that for each particle moving in a certain (tangential) direction in the comoving frame, there must be another particle of the same mass, moving in the opposite (tangential) direction. This configuration was first considered by Einstein, and is known as one with {\em counter rotating} particles. If the particles are further restricted, at least in some spacelike hypersurface, to a region $R_{in} \leq r \leq R_{out}$, where $r$ is the surface area radius in the spherically symmetric space time, we denote the system as an Einstein {\em thick} shell \cite{einstein},  and, in the limit $R_{in} \to  R_{out}$, we have an Einstein {\em thin} shell of counter rotating particles \cite{Hamity}, \cite{Berezin}.

Regardless of the limiting process, we may consider thin shells of counter rotating particles and, as is clear after the appropriate equations are set, still be able to impose further restrictions on the particle motions. These systems, that
are among the simplest dynamical models of matter in spherically symmetric space-times, were studied thirty years ago by Evans \cite{Evans}, who considered a manifold with an embedded time like hypersurface, also with spherical symmetry, that represents the world tube of a
thin shell. All matter content is within the hypersurface so that it is singular and the Darmois-Israel formalism \cite{Israel} is applicable.
In order to get definite equations of motion he also made two important assumptions that we will discuss later: {\em i. particles inside the
shell move along geodesics of the induced metric}, and {\em ii. the absolute value of the angular momentum is the same for all particles.}

Models like that of Evans (as special cases of application of
singular hypersurface theory) have been studied by a number of authors in
different applications, including, for example, the analysis of certain properties
of the evolution of star clusters \cite{Barkov}. The stability of single shells under some general assumptions on the type of matter contents has also been considered in the literature \cite{Lobo} but, as far as we know, the
stability of the trajectories followed by particles within the
shell, or the stability of a shell under possible splitting into separate components have not been analyzed. This is important, because, as we shall
show, the equations of motion are satisfied also for shells that are
intrinsically unstable, and therefore, certain results regarding the
maximum compactness (ratio of radius to mass) may need revision when
stability is taken in consideration. In the case of shells undergoing dynamical evolution, a new feature related to
stability that results from our analysis is the possibility
that, under evolution, an initially stable single shell may split into several separate
components.

The plan of this paper is as follows. In the next Section we review the formulation of the model. In Section III we obtain the equations of motion for a shell with an arbitrary distribution of angular momentum for the constituent particles. Then, in Section IV, we consider some examples of static shells, including a two component shell, that shows explicitly that the equations of motion do not provide un upper bound to the possible angular momentum of the particles. In Section V we analyze single particle orbits, and obtain general conditions for the stability of the shell under single particle ``evaporation''. In Section VI we review the stability of the shells as a whole, finding results in agreement with those of previous authors. Sections VII and VIII are devoted to the analysis of the stability under shell division of both static and dynamic two component shells. Explicit stability conditions that relate the angular momenta to the radius and masses are obtained, and some explicit examples are analyzed. Although restricted to a two component shell, the results should be qualitatively applicable to more general configurations. We also include an Appendix where we show that the thin shell limit of an Einstein shell is a shell with a unique continuous distribution of angular momenta. We close with some additional comments on the results obtained.

\section{ Thin shells dynamics in general relativity}

To construct a general model of a thin, spherically symmetric shell
of matter we consider a spherically symmetric space-time in which
the thin shell corresponds to a timelike
3-surface $\Sigma$ embedded in this space-time. $\Sigma$ is the common boundary of two disjoint manifolds: ${\cal{M}}_I$  and
${\cal{M}}_{II}$. Spherical symmetry implies that
the line elements on ${\cal{M}}_I$ and ${\cal{M}}_{II}$ can be
written as:
\begin{equation}
\label{metII}
ds_{I,II}^2=-F_{I,II}dt_{I,II}^2+\frac{1}{F_{I,II}}dr_{I,II}^2+r_{I,II}^2d\Omega^2
\end{equation}
where $F_{II} = 1-2M_{II}/r_{II}$ and $F_I=1-2M_{I}/r_{I}$, and
$d\Omega= d\theta^2+\sin^2\theta d\phi^2$.

We write
the metric induced on $\Sigma$ as
\begin{equation}\label{metsig}
ds^2_{\Sigma} = d\tau^2 + R(\tau)^2 d\Omega
\end{equation}
where ($\tau$,$\theta$,$\phi$) are coordinates on the shell, with
$\tau$ the proper time of an observer at fixed $(\theta,\phi)$ on
the shell. We take $r_I \leq R(\tau)$, and $r_{II} \geq R(\tau)$. Therefore $\Sigma$ can be described by:
$S_{I,II}(r_{I,II},t_{I,II},\theta,\phi)=r_{I,II}-R(\tau)=0$, and $R(\tau)$ is the radius of the shell.

We will apply the Darmois-Israel formalism for singular hyper
surfaces to obtain the equations of motion of the shell. The non vanishing components of the extrinsic curvatures of $\Sigma$, respectively in ${\cal{M}}_I$ and
${\cal{M}}_{II}$,  in
($\tau$,$\theta$,$\phi$) coordinates are:
\begin{eqnarray}
K^{I,\, II}_{\tau\tau} & = &
 \frac{R^2\ddot{R}+M_{I,\,II}}{R\sqrt{R(R\dot{R}^2+R-2M_{I,\,II})}} \nonumber \\
K^{I,\, II}_{\theta\theta} & = &-\sqrt{R(R\dot{R}^2+R-2M_{I,\,II})} \\
K^{I,\, II}_{\phi\phi} & = &\sin^2 \theta \; K^{I, \,II}_{\theta\theta} \nonumber
\end{eqnarray}
where $\dot{R} = dR/d\tau$, and $\ddot{R} = d^2R/d\tau^2$.

We can write the energy-momentum tensor for a singular hypersurface
$\Sigma$ as:
\begin{equation}
T_{\mu\nu}(x^{\alpha})=\delta(S(x^{\alpha}))S_{\mu \nu}
\end{equation}
Then, Israel's junction conditions \cite{Israel} are
given by:

\begin{equation}
\label{Isr1}
(n^{\alpha}n_{\alpha}) ([K_{ij}]-^3g_{ij}[tr(K)]) = \kappa
S_{\mu\nu}e_i^{\mu}e_j^{\nu}=\kappa S_{ij}
\end{equation}
where for a tensor $A$, we define $[A] = A^{II} -A^{I}$,  $e_i$
stands for $\partial/\partial\tau$, $\partial/\partial\theta$ or
$\partial/\partial\phi$ on the shell, and $n^{\alpha}$ is the unit
normal vector of the shell.

Applying equation (\ref{Isr1}) we have:

\begin{eqnarray}
\label{presg}
\frac{2}{R}\left[\sqrt{\dot{R}^2+1-\frac{2M_{II}}{R}}-\sqrt{\dot{R}^2+1-\frac{2M_{I}}{R}}\right]
& = & \kappa S_{\tau}^{\tau} \\
\label{presg2}
\frac{R^2\ddot{R}+R\dot{R}^2+R-M_{II}}{R\sqrt{R(R\dot{R}^2+R-2M_{II})}}
-\frac{R^2\ddot{R}+R\dot{R}^2+R-M_{I}}{R\sqrt{R(R\dot{R}^2+R-2M_{I})}} & = & \kappa
S_{\theta}^{\theta}=\kappa S_{\phi}^{\phi}
\end{eqnarray}

These equations are completely general, and, once they are
supplemented with an equation of state for the matter contents, they completely
determine the dynamics of the shell. A very simple example is given
by a shell of ``dust'', where, $S_{\theta}^{\theta}= S_{\phi}^{\phi}
=0$. Then, (\ref{presg}) alone, up to integration constants,
determines the dynamics.

In this paper we will be interested in the case where the matter
comprising the shell is made up of non interacting particles, each
particle being subject only to the (average) gravitational field
generated by the rest, and, possibly by some other external source.
In principle, then, one would expect that the particles move along
geodesics of the total gravitational field. We notice, however, that
the Christoffel symbols are discontinuous on the 3-surface. In
particular, a world line contained in the shell that is geodesic
with respect to the connection of region I is not a geodesic for the
connection of region II, and viceversa, and more generally, a geodesic relative to the induced metric on the shell does not correspond to a geodesic in either ${\cal{M}}_I$ or
${\cal{M}}_{II}$. We must, therefore, make more precise what we mean by ``non
interacting particles'' in this context, since we expect the world
lines of such particles to be geodesics in some well defined sense. \cite{Hamity}
We will make the same assumption as Evans \cite{Evans}:
{\em particles inside the shell move along geodesics of the induced
metric}. As a consequence of this assumption, angular momentum is a
constant of the motion for each particle. Moreover, the restriction
to spherically symmetric configurations, implies that for every
particle moving with a certain velocity in a certain direction
relative to the shell we must have another moving in the opposite
direction, with the same speed, i.e., ``counter - rotating" particles, as in the Einstein model
\cite{einstein}, or, for the dynamical case, as in \cite{Evans}. We notice that this assumption does not restrict the possible values of the particle's angular momentum to a single value. A particular example of a shell with a continuous distribution of values of the particle's angular momenta, is given by the limit of shells of counter rotating
particles with non vanishing thickness (Einstein shells) as the
thickness goes to zero. This is discussed in more detail in the
Appendix, where we show that the limit thin shell must contain particles with a range of values of angular momentum, so that, in particular, the model analyzed by Evans does not correspond to the limit of an Einstein shell. It is, nevertheless, expected that different limiting distributions, either discrete or continuous, would result if instead of an Einstein shell one considers more general shells of Vlasov type matter. \cite{AndreasBuch}

Going back to the general formulation of the problem, and taking into account the symmetries of the problem, the 4-velocity of
a particle in the shell is completely determined by the modulus of
its angular momentum per unit rest mass $L$ and the angle
$\chi=\tan^{-1}(u_{\theta}/u_{\phi})$, \cite{Evans}, and may be written in the form:

\begin{equation}
\label{u}
u^{\mu}(\tau,\theta,\phi)=\left[\left(1+L^2/R(\tau)^2\right)^{1/2},0,
\frac{L}{R(\tau)^2}\sin(\chi),\frac{L}{R(\tau)^2}\frac{\cos(\chi)}{\sin(\theta)}\right]
\end{equation}
where $\mu = \tau,n,\theta,\phi$, with $n$ a Gaussian normal coordinate with respect to the shell.
Therefore, we can write the surface stress-energy tensor as:

\begin{equation}
\label{Sthin} S^{ij}(\tau)=\frac{1}{2\pi}\int \sigma(L,\tau)
u^{i}(\chi,L)u^{j}(\chi,L)d\chi dL
\end{equation}
where $i = \tau, \theta, \phi$, and $\sigma(L,\tau)$ is the proper
mass density of particles with modulus of angular momentum per unit
rest mass $L$, at time $\tau$. As we indicate in the next Section,
particle number conservation leads to a more explicit expression for
the functional form of $\sigma(L,\tau)$.

\section{ Equations of motion}

The treatment in \cite{Evans} was restricted to particles with a
single value of $L$. This considerably simplifies the expression for
$S^{ij}$ and, as a consequence, the form of the equations of motion
for the shell. Here we consider the more general case, where the
angular momentum of the particles can take arbitrary values.

We can use now particle number conservation to obtain first integrals for
the Einstein equations. Equation (3.17) of \cite{ComKatz} implies that
the integral on a  2-surface contained in the shell world tube defined by $\tau=\tau_0$ :
\begin{equation}
\frac{1}{\mu} \int_{\tau=\tau_0} \sigma (L,\tau) d\Sigma_{\alpha} u^{\alpha} dL
\end{equation}
where $\mu$ is the proper mass of the individual particles, is
independent of the choice of $\tau_0$ and represents the total
number of particles inside the shell. Moreover, because of the lack
of interaction between the shell particles, the number density of
particles whose angular momentum is $L$:
\begin{eqnarray}
\label{nL}
 n(L)= \frac{1}{\mu} \int_{\tau=\tau_0} \sigma (L,\tau)d\Sigma_{\mu} u^{\mu}  &=&
 \frac{1}{\mu}\int \sigma(L,\tau_0)\left(1+\frac{L^2}{R(\tau_0)^2}\right)^{1/2}
 R(\tau_0)^2 \sin(\theta)d\theta d\phi  \nonumber \\
&=&\frac{4\pi}{\mu}\sigma(L,\tau_0) R(\tau_0)\sqrt{L^2+R(\tau_0)^2}
\end{eqnarray}
is also constant. Taking this into account, the components $S_i{}^j$ can
be written:
\begin{equation}
\label{S}
 S_i{}^j=\frac{\mu}{4\pi R^3} {\mbox{ diag}} \left[-\int n(L) \sqrt{R^2+L^2} dL,
 \int \frac{n(L) L^2}{2\sqrt{L^2+R^2}} dL,\int \frac{n(L) L^2}{2\sqrt{L^2+R^2}} dL\right]
\end{equation}

A simple consequence of (\ref{S}) is that,
\begin{equation}\label{S1}
- S_{\tau}{}^{\tau}-2 S_{\theta}{}^{\theta} = \frac{\mu}{4\pi R}\int
\frac{n(L) }{\sqrt{L^2+R^2}} dL \geq 0
\end{equation}
and, therefore,
\begin{equation}\label{S2}
- S_{\tau}{}^{\tau}\geq 2 S_{\theta}{}^{\theta}
\end{equation}
with equality approached only in the limit $L^2 \to \infty$,
corresponding to ultra relativistic (or massless) particles. If we
define $\alpha = S_{\theta}{}^{\theta}/(-S_{\tau}{}^{\tau})$, then,
\begin{equation}\label{alfa1}
\alpha = \frac{1}{\int n(L) \sqrt{R^2+L^2} dL}\int \frac{n(L) L^2}{2\sqrt{L^2+R^2}} dL,
\end{equation}
and equation (\ref{S2}) implies $ 0 \leq \alpha \leq 1/2$, and from
(\ref{presg2}), we have,
\begin{equation}\label{S3}
\ddot{R} = -\frac{1}{2 R} \left[ 1+(\dot{R})^2 -(4
\alpha+1)\sqrt{1-\frac{2 M_{I}}{R}+(\dot{R})^2}\sqrt{1-\frac{2
M_{II}}{R}+(\dot{R})^2}\right]
\end{equation}
Notice that, given $n(L)$, $\alpha$ depends only on $R$, and,
therefore, (\ref{S3}) is an integro differential equation for $R(\tau)$.

Equation (\ref{S}) together with (\ref{presg}) can be used to obtain:
\begin{equation}
\label{1stint}
 \frac{2}{R}\left[\sqrt{\dot{R}^2+1-\frac{2M_{II}}{R}}
 -\sqrt{\dot{R}^2+1-\frac{2M_{I}}{R}}\right] = -\frac{2\mu}{R^3} \int n(L) \sqrt{R^2+L^2} dL
\end{equation}
and it can be checked that (\ref{1stint}) is a first integral of (\ref{S3}).

Defining $f(R)=\int n(L)
\sqrt{R^2+L^2} dL$, this expression can be written as ($C=2\mu$):
\begin{equation}
\label{eqmovg}
 \dot{R}^2 = V(R) = -1+\frac{M_{II}+M_{I}}{R} + \frac{C^2 f(R)^2}{16R^4}+\frac{4(M_{II}-M_{I})^2R^2}{C^2f(R)^2}
\end{equation}

Equation (\ref{eqmovg}) is a generalization of equation (2.35) in
\cite{Evans} for an arbitrary angular momentum distribution $n(L)$.
We can see that the qualitative features of the motion of the shell
are the same as those described by Evans. $V(R)$, $f(R)$ and its
first derivative have the following asymptotic behavior:

\begin{equation}\label{asymp1}
R \rightarrow 0 \Rightarrow
\begin{cases}
f(R) \rightarrow N \langle L\rangle\\
f^{'}(R) \rightarrow 0 \\
V(R) \rightarrow \frac{C^2 N^2\langle L\rangle^2}{16R^4}
\end{cases}
\end{equation}
and,
\begin{equation}\label{asymp2}
R \rightarrow \infty \Rightarrow
\begin{cases}
f(R) \rightarrow  R N  \\
f^{'}(R) \rightarrow N \\
V(R) \rightarrow -1 + \frac{4(M_{II}-M_{I})^2}{C^2N^2}
\end{cases}
\end{equation}
where $N$ is the total number of particles and $\langle L\rangle$ is
the mean value of $L$. The ``binding energy per particle'' is the
constant $-1+4(M_{II}-M_{I})^2/(C^2N^2)$. If $2(M_{II}-M_{I}) > CN$ the binding energy is
positive and the shell can reach infinity, otherwise, the shell may only reach
a maximum possible radius. There are also solutions where the shell has a lower turning point, that is, the radius reaches a minimum value, which can be obtained solving $V(R)=0$, for $R$. Some care should be exercised here, however, because (\ref{eqmovg}) does not take automatically into account the signs of the square roots in (\ref{1stint}). In the following Sections we consider some simple applications of this formalism, paying particular attention to the dynamical stability of the resulting models.

\section{Static shells}

The simplest dynamics results, of course, if the shell is static,
with a fixed radius $R=R_0$. In this case we must have both
$\dot{R}=0$, and $\ddot{R}=0$. From (\ref{eqmovg}), this implies
both $V|_{R=R_0}=0$, and $(dV/dR)|_{R=R_0}=0$, and, therefore, the
static shell radius corresponds to a critical point of $V(R)$.  These conditions are equivalent to setting $\dot{R}=0$, and $\ddot{R}=0$ in (\ref{presg}) and (\ref{presg2}), i.e.,
\begin{eqnarray}
\label{presgsta}
\frac{2}{R}\left[\sqrt{ 1-\frac{2M_{II}}{R}}-\sqrt{ 1-\frac{2M_{I}}{R}}\right]
& = & -\frac{2\mu}{R^3} \int n(L) \sqrt{R^2+L^2} dL \\
\label{presgsta2}
\frac{ R-M_{II}}{R\sqrt{R( R-2M_{II})}}
-\frac{ R-M_{I}}{R\sqrt{R(R-2M_{I})}} & = & \frac{2\mu}{R^3} \int \frac{n(L)L^2}{ \sqrt{R^2+L^2}} dL
\end{eqnarray}

For a given distribution function $n(L)$ these equations relate the equilibrium radius $R_0$, and the masses $M_{I}$, and $M_{II}$ with the parameters that characterize $n(L)$. Some simple examples considered below illustrate this point.

\subsection{Single Component Shell}
Static single component shells are a particular instance of the
systems analyzed in \cite{Evans}. In this case we have $n(L)=N
\delta (L-L_0)$, that is, all particles have the same angular
momentum $L_0$. The resulting equations are:
\begin{eqnarray}
\frac{2}{R}\left[\sqrt{1-\frac{2M_{II}}{R}}-\sqrt{1-\frac{2M_{I}}{R}}
\right] &=& - \frac{CN\sqrt{R^2+L_0^2}}{R^3} \\
\frac{R-M_{II}}{R\sqrt{R(R-2M_{II})}}
-\frac{R-M_{I}}{R\sqrt{R(R-2M_{I})}} & = & \frac{CNL_0^2}{2
R^3\sqrt{R^2+L_0^2}}
\end{eqnarray}

It is interesting to solve these equations for $L_0^2$ and $CN$.
We find,
\begin{equation}\label{Lcrit}
L_0^2 =  \frac{(R-\sqrt{R-2 M_{I}}\sqrt{R-2M_{II}})R^2}{3\sqrt{R-2 M_{I}}\sqrt{R-2M_{II}}-R}
\end{equation}
For reasons that will be made clear below, we shall call this expression for $L_0$, the {\em critical} value of the angular momentum for a shell
of radius $R$, with exterior mass $M_{II}$, and interior mass $M_{I}$. For $CN$ we get,
\begin{equation}\label{CN1}
CN = \frac{(\sqrt{R-2 M_{I}}-\sqrt{R-2 M_{II}})R\sqrt{6
\sqrt{R-2 M_{I}}\sqrt{R-2M_{II}}-2 R}}{\sqrt{R-2 M_{I}}\sqrt{R-2M_{II}}}
\end{equation}
and we must require $M_{II} \geq M_{I} \geq 0$, so that $CN \geq 0$. We
see that meaningful values of $L_0^2$ and $CN$ are only obtained
provided $3 \sqrt{R-2 M_{I}}\sqrt{R-2M_{II}} > R$, and this implies
\begin{equation}\label{bound}
\frac{R}{M_{II}} > \frac{9}{4}\left[\frac{1}{2} +
\frac{\beta}{2}+\frac{1}{6} \sqrt{9-14 \beta +9 \beta^2}\right]
\geq \frac{9}{4}
\end{equation}
where $\beta = M_{I}/M_{II}$ and the lower bound on the right hand
side of (\ref{bound}) corresponds to $\beta=0$. In this case
($M_{I}=0$), we also have,
\begin{equation}\label{Lc}
L_0^2= \frac{R^2}{4R-9M_{II}}\left[3
M_{II}-R+\sqrt{R(R-2M_{II})}\right]
\end{equation}

The bound given by (\ref{bound}) has been considered as the correct
lower bound for the compactness (i.e. ratio $R/M_{II}$) for this
type of thin shells, with implications, for instance, for the
maximum gravitational red shift for light emitted from their
surface \cite{AndreasBuch}. However, as we shall prove in the following sections in our analysis
of the stability of the shells, this bound cannot
be approached if the shell dynamics is governed only by
gravitational interactions.

\subsection{Shells with two components}

A detailed analysis of the dynamics of shells containing particles with more than one value of $L$ is rather complex in general. Nevertheless, even if we restrict to two
components, that is, to two values of $L$, some interesting features
appear that deserve closer examination. In this case we have that the
number density for particles with angular momentum $L$
can be written as $n(L)=N_1\delta(L-L_1)+N_2\delta(L-L_2)$, and,
therefore, after some simplifications, we have,
\begin{eqnarray}
\label{29}
 \sqrt{1-\frac{2M_{I}}{R}}-\sqrt{1-\frac{2M_{II}}{R}} &=&  \frac{C N_1\sqrt{R^2+L_1^2}}{2R^2}+ \frac{C N_2\sqrt{R^2+L_2^2}}{2 R^2} \\
\label{30}
\frac{R-M_{II} }{\sqrt{R(R-2M_{II} )}} - \frac{R-M_{I} }{\sqrt{R(R-2M_{I} )}} & = & \frac{C N_1 L_1^2}{2
R^2\sqrt{R^2+L_1^2}}+\frac{CN_2 L_2^2}{2 R^2\sqrt{R^2+L_2^2}}
\end{eqnarray}

We may solve these equations for $N_1$, and $N_2$,
\begin{eqnarray}
\label{sta02}
 C N_2 & =&
\frac {2 R^2 (L_0^2-L_1^2) \sqrt {{R}^{2}+{L_2}^{2}}} {(R^2+L_0^2)(L_2^2-L_1^2)}
\left( \sqrt {1-\frac{2 M_{I}}{R}}- \sqrt {1-\frac{2 M_{II}}{R}}\right)  \nonumber \\
C N_1& = &
\frac {2 R^2 (L_2^2-L_0^2) \sqrt {{R}^{2}+{L_1}^{2}}} {(R^2+L_0^2)(L_2^2-L_1^2)}
\left( \sqrt {1-\frac{2 M_{I}}{R}}- \sqrt {1-\frac{2 M_{II}}{R}}\right)
\end{eqnarray}
where $L_0$ is the critical value of $L$, given by (\ref{Lcrit}). This
result implies that $N_1$ and $N_2$ will be positive, as required by
the particle interpretation,  provided only that either $L_2 > L_0 >
L_1$, or $L_1 > L_0 > L_2$, that is, that one the $L_i$'s is larger
and the other smaller than the critical value $L_0$, for the given
values of $R$ and $M$. In fact, if we consider a static shell with any number of components $L_i$, it is not difficult to show that if the corresponding $N_i$ are positive, then there must be values both larger and smaller than $L_0$ among the $L_i$. \\

This can be seen as follows. For a static shell with $n$ components, we have $N_i, i =1\dots n$, and corresponding $L_i, i= 1 \dots n$, and we may assume $L_1\leq L_2\leq \dots \leq L_n$. Then from the equations generalizing (\ref{29}), and (\ref{30}), and the definition of $L_0$, we have,
\begin{eqnarray}
\label{teo1}
C N \sqrt{R^2+L_0^2} &=&  \sum_{i=1}^n C N_i \sqrt{R^2+L_i^2}   \\
\label{teo2}
  C N \frac{L_0^2}{\sqrt{R^2+L_0^2}} &=&  \sum_{i=1}^n C N_i \frac{L_i^2}{\sqrt{R^2+L_i^2}}
\end{eqnarray}
Combining (\ref{teo1}), and (\ref{teo1}), we have,
\begin{equation}
\label{teo3}
  C N \frac{R^2}{\sqrt{R^2+L_0^2}} =  \sum_{i=1}^n C N_i \frac{R^2}{\sqrt{R^2+L_i^2}}
\end{equation}
then, since all $N_i \geq 0$, if we assume that $L_1 \geq L_0$, from (\ref{teo1} we have,
\begin{equation}
\label{teo1a}
C N \sqrt{R^2+L_0^2} \geq  \sqrt{R^2+L_0^2} \sum_{i=1}^n C N_i
\end{equation}
while from (\ref{teo3}) we obtain,
\begin{equation}
\label{teo2a}
  C N \frac{R^2}{\sqrt{R^2+L_0^2}} \leq  \frac{R^2}{\sqrt{R^2+L_0^2}} \sum_{i=1}^n C N_i
\end{equation}
which contradicts (\ref{teo1a}), and, therefore, the assumption that $L_1 \geq L_0$, and $N_i > 0$. Similarly, we may prove that we must have $L_n \geq L_0$, with the equal sign holding only if $L_i=L_0$ for all $i$.

Going back now to the two component shell, we find that (\ref{sta02}) is a very intriguing result, because
it implies that the angular momentum of the particles in one of the
components, say $L_2$, can take arbitrarily large values for any
admissible $R$ and $M$, and we still have an admissible solution. But this is also true even if $R >> M$,
where the regime is essentially Newtonian, and since large $L_2$
implies large speeds for the particles, it is hard to see how the
particles would remain bound to the shell, if only the gravitational
attraction of the other particles is responsible for the particle
trajectory. Similar problems arise when we consider more than two
components. As a first step towards understanding this property of
the solutions of the equations of motion of the shell, in the next
Section we analyze the stability of single particle orbits.

\section{Stability of single particle orbits}

We recall that the hypersurface $\Sigma$ containing the shell may be considered as a boundary for either ${\cal{M}}_{I}$ or ${\cal{M}}_{II}$.
Let us fix now our attention on one of the particles moving on the shell with 4-velocity $u^{\alpha}$. From metric continuity, we may consider
this particle as contained in either ${\cal{M}}_{I}$ or ${\cal{M}}_{II}$, with the same 4-velocity $u^{\alpha}$. In general, from the point
of view of ${\cal{M}}_{I}$ or ${\cal{M}}_{II}$, namely considering its world line as a curve on either part of the space time, the particle does
not follow a geodesic path, but rather, it has a non vanishing 4-acceleration. To analyze the stability of the orbit of a single particle moving
on the shell with 4-velocity $u^{\alpha}$, we consider an infinitesimal (radial) displacement of its world line, so that the particle may now be
considered as moving freely in either ${\cal{M}}_{I}$ or ${\cal{M}}_{II}$, depending on the direction of the displacement. Because it moves now
along a geodesic close, but outside $\Sigma$, the particle will acquire an acceleration relative to the shell. The motion will therefore be {\em
stable} if the acceleration points {\em towards} the shell, but it will be {\em unstable} is the relative acceleration points {\em away} from
the shell. To illustrate this point, and before we discuss the general case, let us consider a static shell, of radius $R$ and (external) mass
$M_{II}$, which, in accordance with our previous discussion may contain particles with arbitrarily large angular momentum $L_p$. The metric in
the external region ${\cal{M}}_{II}$, ($r > R$) may be written as,
\begin{equation}\label{singl01}
    ds^2= -(1-2 M_{II}/r) dt^2 + (1-2M_{II}/r)^{-1} dr^2 + dr^2 d \Omega^2
\end{equation}
Assuming equatorial motion, ($\theta=\pi/2$), and calling $\sigma$ the proper time along the particle world line, the geodesic equations imply $
d\phi/d\sigma=L_p/r^2$,
\begin{equation}
\label{singl02}
  \left(\frac{dt}{d\sigma}\right)^2 = \frac{r^2}{(r-2 M_{II})^2} \left(\frac{dr}{d\sigma}\right)^2 +\frac{L_p^2+r^2}{r(r-2M_{II})}
\end{equation}
and,
\begin{equation}
\label{singl03}
  \frac{d^2r}{d\sigma^2} = -\frac{M_{II}}{r^2}+\frac{r-3M_{II}}{r^4}L_p^2
\end{equation}
Since, for a particle on the shell ($r=R$) we have $dr/d\sigma=0$, for $R>3M_{II}$ the motion will be stable against "single particle evaporation", only
if, for $R > 3M_{II}$, we have $L_P^2 < M_{II} R^2/(R-3M_{II})$, while for $R \leq 3M_{II}$, all orbits are stable. This is, of course, a consequence of the fact, that in
Schwarzschild's space time, time like geodesics cannot have a lower turning for $r \leq 3M$. We also remark that for a single component shell, $L_p=L_0$, and, as can be checked, the shell is always stable under single particle evaporation.

To analyze the more general case of a dynamic shell it will be advantageous to introduce a coordinate system $(\tau, \rho,\theta,\phi)$, adapted
to the shell, where $\rho$ is a gaussian coordinate normal to $\Sigma$. We are only interested in the limit $\rho \to 0$. Then, keeping only
terms up to linear order in $\rho$, in the neighbourhood of the shell ($\rho=0$), we may write the metric in the form,
\begin{equation}\label{gaus1}
ds^2= -(1+A \rho) d\tau^2 +2 B\rho d\tau d\rho +(1+C \rho ) d\rho^2 +(R^2+ D \rho ) d\Omega^2 + {\cal{O}}(\rho^2)
\end{equation}
where $A$, $B$, $C$ and $D$ are functions of $\tau$ given,
\begin{eqnarray}
\label{gaus2}
  A &=& \frac{2 ( R^2\ddot{R} +M_{\pm})}{R \sqrt{(R
\dot{R}^2 +R- 2M_{\pm})R}} \nonumber \\
  B &=& \frac{(6M_{\pm} -3 R -2 R \dot{R}^2)M_{\pm} \dot{R}}{R(R-2M_{\pm})^2} \nonumber  \\
  C &=& \frac{2 M_{\pm} (2M_{\pm}-R-2 R \dot{R}^2) \sqrt{R(R\dot{R}^2
+R-2M_{\pm})}}{R^2(R-2M_{\pm})^2} \\
  D &=& 2\sqrt{(R \dot{R}^2 +R- 2M_{\pm})R} \nonumber
\end{eqnarray}
where $M_{\pm}=M_{II}$ for $\rho >0$ (the exterior region), and $M_{\pm}=M_{I}$ for $\rho < 0$ (the interior region). $R=R(\tau)$ is the shell
radius, and dots indicate derivatives with respect to $\tau$. We also require $M_{II} > M_{I}$, for consistency. This metric is continuous, with
discontinuous first derivatives. For a particle instantaneously moving on the shell we have $d \rho/d\eta =0$, where $\eta$ is the proper time
along the particle world line. Considering again an infinitesimal (positive) radial displacement, so that the particle follows a geodesic path in
${\cal{M}}_{II}$, the motion will be stable only if for the resulting world line $d^2 \rho/d \eta^2 < 0$. Without loss of generality we may
consider equatorial motion with $\theta= \pi/2$. Then we have,
\begin{eqnarray}
\label{gaus3a}
  u^{\phi} &=&  \frac{L_p}{R^2}  \nonumber \\
  \left(u^{\tau}\right)^2 &=& 1+\frac{L_p^2}{R^2}
\end{eqnarray}
where $u^{\alpha} = d x^{\alpha}/d\eta$. Then, the geodesic equations of motion imply,
\begin{equation}\label{gaus4}
\frac{ d^2 \rho}{d \eta^2}=-\frac{M_{II}+R^2\ddot{R}}
{R\sqrt{R(R\dot{R}^2+R-2M_{II})}}+L^2\frac{R-3M_{II}-R^2\ddot{R}+R\dot{R}^2}{R^3\sqrt{R(R\dot{R}^2+R-2M_{II})}}
\end{equation}
This results reduces to (\ref{singl03}) for the static case $\ddot{R}=0$, $\dot{R}=0$, the apparent difference being due to the different
definitions of $r$ and $\rho$. For $R > 3 M_{II}$ and $\ddot{R}\leq 0$, this result implies that for sufficiently large $L_p$, we have  $d^2
\rho/ d \eta^2 > 0$, and the resulting motion is unstable. The, perhaps, unexpected result is that for sufficiently large $\ddot{R} > 0$, that
is, a shell undergoing fast accelerated expansion, we may have $d^2 \rho/ d \eta^2 < 0$ for all $L_p$, so that the shell is stable against
``single particle evaporation". One can check that this condition is consistent with the assumption that the shell is made out of counter
rotating particles, but we shall not give the details here.

We may also consider a negative infinitesimal displacement. In this case, we obtain again (\ref{gaus4}), but now with $M_{I}$, replacing
$M_{II}$, and now stability corresponds to $d^2 \rho/ d \eta^2 < 0$. The analysis is similar to that for (\ref{gaus4}), and will not be repeated
here.

The foregoing considerations apply only to what we might call `` single particle evaporation''. Perhaps of much greater interest is the
stability of the shell as a whole. Here, however, we must distinguish different modes. We shall restrict to modes that preserve the spherical
symmetry, and consider two cases. The first, analyzed in the next Section, will be the stability of stationary configurations, and the second,
in the case of shells with a distribution of values of $L$, the stability under separation of components.

\section{Stability of static solutions}

To analyze the stability of static solutions we consider again (\ref{S3}). Setting now $\ddot{R} =\dot{R} = 0$, we find,
\begin{equation}\label{S3a}
0 = -\frac{1}{2 R_0} \left[ 1  -(4 \alpha(R_0)+1)\sqrt{1-\frac{2 M_{I}}{R_0} }\sqrt{1-\frac{2 M_{II}}{R_0} }\right]
\end{equation}
where $R_0$ is the static shell radius. This can be solved for $\alpha$,
\begin{equation}\label{S3b}
\alpha(R_0) = \frac{R_0 -\sqrt{R_0- 2 M_{I} }\sqrt{R_0-2 M_{II} }}{4\sqrt{R_0- 2 M_{I} }\sqrt{R_0-2 M_{II} }}
\end{equation}
Since $\alpha(R_0) \leq 1/2$, we obtain the bound \cite{AndreasBuch},
\begin{equation}\label{bigR6}
 R_0     \geq  \frac{9}{8} \left(M_{I}+M_{II}\right) +\frac{3}{8}\sqrt{9\left(M_{I}-M_{II}\right)^2+4 M_{I}M_{II}} \geq \frac{9}{4}M_{II}
\end{equation}
for any shell of counter rotating particles. The lowest value can only be approached for $M_{I}=0$. This is the same as (\ref{bound}), but in this case there is no restriction on $n(L)$.

We consider now a small
perturbation of the shell radius that preserves spherical symmetry and the angular momentum distribution $n(L)$. Then $\alpha$ may be
considered as a function of $R$ only, and we have $\alpha'=d \alpha/dR < 0$. We introduce $\delta R(t)$ such that $R =R_0 + \delta R(t)$, and
expand (\ref{S3}) to first order in $\delta R$ and its time derivatives. After some simplifications using also (\ref{S3b}), we find,
\begin{equation}\label{Sta01}
\frac{d^2 \delta R}{dt^2} =\frac{2 \sqrt{R_0-2M_{I}}\sqrt{R_0-2M_{II}}}{R_0^2}\ \alpha'(R_0)\delta R +\frac{(M_{I}+M_{II})R_0-4M_{I}M_{II}}{2
R_0^2 (R_0-2 M_{I})(R_0-2M_{II})} \delta R
\end{equation}

Since we take $M_{II} > M_{I}$, and we must have $R_0 > 2 M_{II}$, the first term on the right hand side
of (\ref{Sta01}) is always negative, contributing to stability, while the second term is always positive,
tending to make the system unstable.

We may consider now two limits. First, from (\ref{alfa1}), for large $R_0$, assuming $R_0>>L$ in (\ref{alfa1}), we have,
\begin{eqnarray}\label{bigR1}
\alpha & \simeq  & \frac{\int{n(L)L^2 dL}}{2 R_0^2\int{n(L)dL}} = \frac{1}{2R_0^2} <L^2> \nonumber \\
\alpha'(R_0) & \simeq & - \frac{\int{n(L)L^2 dL}}{ R_0^3\int{n(L)dL}} = - \frac{1}{R_0^3} <L^2>
\end{eqnarray}
where angle brackets indicate average with respect to $n(L)$. Then, from the the staticity condition (\ref{S3b}), we have,
\begin{equation}\label{bigR2}
    <L^2>\simeq \frac{1}{2} \left(M_{I}+M_{II}\right) R_0
\end{equation}
and, replacing in (\ref{Sta01}), we find,
\begin{equation}\label{bigR3}
\frac{d^2 \delta R}{dt^2} \simeq -\frac{(M_{I}+M_{II})}{2 R_0^3} \delta R
\end{equation}
Therefore, all static shells are stable (as regards the mode considered here) for sufficiently large $R_0$.

The other limit is as $R_0$ approaches the minimum value allowed by (\ref{S3b}) and $\alpha \leq 1/2$. But this requires $\alpha \simeq 1/2$, which in turn requires large values of $L$ as compared with $R_0$ in (\ref{alfa1}). In this case we have,
\begin{eqnarray}\label{bigR4}
\alpha & \simeq  & \frac{1}{2}-  \frac{R_0^2}{2 \int{n(L)L dL}} \left[\int{n(L)\frac{1}{L} dL}\right]=\frac{1}{2}- \frac{R_0^2}{2} \frac{<L^{-1}>}{<L>} \nonumber \\
\alpha'(R_0) & \simeq &   - \frac{R_0^2}{\int{n(L)L dL}}\left[ \int{n(L)\frac{1}{L} dL}\right]=-  R_0 \frac{<L^{-1}>}{<L> }
\end{eqnarray}
Then, as $R_0$ approaches the minimum value, we have $\alpha'(R_0) \to 0$, and, therefore, from  (\ref{Sta01}), we conclude that {\em all} shells are unstable in this limit. To find the minimum {\em stable} radius $R_s$, from (\ref{Sta01}), we see that for stability we must have,
\begin{equation}\label{bigR7}
\alpha'(R_s) \leq -\frac{\left(M_{I}+M_{II}\right) R_s -4 M_{I}M_{II}}{4 (R_s-2M_{I})^{3/2}(R_s-2M_{II})^{3/2}}
\end{equation}
from which we may obtain the minimum equilibrium radius $R_s$, for a given distribution function $n(L)$. As an example, for a single component shell we have $\alpha(R_s) = L_0^2/(2(L_0^2+R_s^2))$, and (\ref{S3b}) implies,
\begin{equation}\label{bigR8}
L_0^2 =\frac{R_s^2\left(R_s-\sqrt{R_s-2M_{I}}\sqrt{R_s-2M_{II}}\right)}{3\sqrt{R_s-2M_{I}}\sqrt{R_s-2M_{II}}-R_s}
\end{equation}
Then, from (\ref{bigR7}) we find that $R_s$ is a solution of,
\begin{equation}\label{bigR9}
\frac{ 4{R_s}^{2}-9 \left( M_{I}+M_{II}
 \right) R_s+20 M_{II}\,M_{I}}
 {2 \sqrt {R_s-2
 M_{I}}\sqrt {R_s-2 M_{II}}}
 - \frac{  2 {R_s}^{2}-3
 \left( M_{I}+M_{II} \right) R_s+6 M_{II} M_{I}
 }{R_s} =0
\end{equation}
The appropriate root of (\ref{bigR9}) grows smoothly as $M_{I}$ grows from $M_{I}=0$ up to $M_{I}=M_{II}$. Near those extremes it  satisfies,
\begin{equation}\label{bigR10}
R_s \simeq \left(\frac{51}{16}+\frac{3\sqrt{33}}{16}\right)M_{II} + \left(\frac{3}{16}+\frac{19\sqrt{33}}{528}\right) M_{I} \simeq 4.26 M_{II}+0.39 M_{I}
\end{equation}
for $M_{I}\to 0$, and,
\begin{equation}\label{bigR11}
R_s \simeq  6 M_{II} - 3 \left(M_{II}-M_{I}\right)
\end{equation}
for $M_{I} \to M_{II}$. Therefore, the lower bound on $R/M_{II}$ set by the stability criterion is almost twice as large as that given by (\ref{bound}). (See also \cite{Lobo} for related results.)

\section{ Stability under shell division of static two component shells.}

As indicated in Section IV, in the case of shells with a distribution of values of $L$, one can have solutions of the equations of motion even if the shell contains particles with arbitrarily large values of $L$. As shown in Section V, this may lead to instabilities under ``single particle evaporation'', but it is natural to question the possible stability of such configurations as a whole. It turns out that the general problem is rather cumbersome and hard to handle. Nevertheless, one can get important information and insight by considering first static shells. In this Section we consider the stability of such shells
under the further, but not trivial, simplification of assuming only two components, one with angular momentum $L_1$, and other with $L_2$, with $L_2 > L_1$. This configuration was already analyzed in Section IV, to obtain the static solution, assuming an inner mass $M_{I}$, outer mass $M_{II}$, and shell radius $R_0$. The central idea here will be to assume that the $L_1$ and $L_2$ components are infinitesimally displaced from the static radius $R_0$, the first to the inside and the second to the outside, so that we have now an intermediate vacuum region separating the two resulting shells. Let $M_{int}$ be the mass parameter for this intermediate region. On account of the derivations in Section III, and assuming that the shells acquire no velocity in the displacement, we have,
\begin{eqnarray}\label{statwo1a}
\frac{d^2 R_1}{d \tau_1^2}& = & -\frac{1}{2R_0}\left[1-\left(\frac{2 L_1^2}{R_0^2+L_1^2}+1\right)\sqrt{1-\frac{2M_{int}}{R_0}}\sqrt{1-\frac{2M_{I}}{R_0}}\right]  \\
\label{statwo1b}
\frac{d^2 R_2}{d \tau_2^2}& = & -\frac{1}{2R_0}\left[1-\left(\frac{2 L_2^2}{R_0^2+L_2^2}+1\right)\sqrt{1-\frac{2M_{int}}{R_0}}\sqrt{1-\frac{2M_{II}}{R_0}}\right]
\end{eqnarray}
where $R_1$ and $R_2$ are the shell radii and $\tau_1$ and $\tau_2$ the corresponding proper times. We also have,
\begin{eqnarray}
\label{statwo2a}
\sqrt{1-\frac{2M_{int}}{R_0}} &=& \sqrt{1-\frac{2M_{I}}{R_0}} -\frac{ C N_1 \sqrt{R_0^2+L_1^2}}{2 R_0^2}    \\
\label{statwo2b}
\sqrt{1-\frac{2M_{int}}{R_0}} &=& \frac{ C N_2 \sqrt{R_0^2+L_2^2}}{2 R_0^2}- \sqrt{1-\frac{2M_{II}}{R_0}}
\end{eqnarray}
We may use now (\ref{sta02}) to eliminate $CN_1$ and $CN_2$ from these equations and solve for $M_{int}$. We find,
\begin{eqnarray}\label{statwo3}
\sqrt{1-\frac{2M_{int}}{R_0}} & = &
\frac{(L_0^2-L_1^2)(R_0^2+L_2^2)}{(L_2^2-L_1^2)(R_0^2+L_0^2)}\sqrt{1-\frac{2M_{I}}{R_0}} \nonumber \\
& & +
\frac{(L_2^2-L_0^2)(R_0^2+L_1^2)}{(L_2^2-L_1^2)(R_0^2+L_0^2)}\sqrt{1-\frac{2M_{II}}{R_0}}
\end{eqnarray}
which is of the form,
\begin{equation}\label{statwo4}
\sqrt{1-\frac{2M_{int}}{R_0}} =
w_1\sqrt{1-\frac{2M_{I}}{R_0}}  + w_2 \sqrt{1-\frac{2M_{II}}{R_0}}
\end{equation}
with $w_i >0$ and $w_1+w_2=1$, so that we always have $M_{I} \leq M_{int} \leq M_{II}$, as expected.

From these equations we immediately conclude that the shell will be stable under separation into two single component shells only if $d^2 R_1/d \tau_1^2 > 0$ and $d^2 R_2/d \tau_2^2 <0$, but unstable otherwise. The critical condition is then $d^2 R_i/d \tau_i^2 = 0$. It is interesting that both conditions are satisfied if $L_1$ and $L_2$ are related by,
\begin{equation}\label{statwo5}
\frac{(3 L_2^2+R_0^2)\sqrt{R_0-2M_{II}}}{(R_0^2+L_2^2)} =\frac {(3 L_1^2+R_0^2)\sqrt{R_0-2M_{I}}}{(R_0^2+L_1^2)}
\end{equation}

We can check that this is indeed a critical condition by computing, e.g., the partial derivative of the R.H.S. of (\ref{statwo1b}) with respect to $L_2$, keeping $L_1$, $R_0$, $M_{I}$, and $M_{II}$ fixed, evaluated assuming (\ref{statwo5}) holds. The result is,
\begin{equation}\label{statwo6}
{\frac  {4 L_2\,{R_0}^{2} \left( {L_2}^2-{L_0}^{2} \right)  \left( R_0-2\,M_{II} \right) \sqrt {R_0-2\,{
M_{I}}}}{ \left( \sqrt {R_0-2\,M_{I}}-\sqrt {R_0-2\,{
M_{II}}} \right)  \left( 3\,{L_2}^{2}+{R_0}^{2} \right)
 \left( {R_0}^{2}+{L_2}^{2} \right) ^{2} \left( {R_0}^{
2}+{L_0}^{2} \right) }} > 0
\end{equation}
Therefore, the shell becomes unstable for values of $L_2$ larger than the one satisfying (\ref{statwo5}). Similar results hold for $L_1$, except that here the shell becomes unstable for values of $L_1$ {\em less} than the critical one given by (\ref{statwo5}).

The relation (\ref{statwo5}) must also comply with the conditions $L_1^2 \leq L_0^2$, and  $L_0^2 \leq L_2^2$. We can see from (\ref{statwo5}) that, for $L^2_i >0$, $L_1^2$ and $L_2^2$ are monotonic functions of each other. Then, the lowest possible value for stability for , e.g., $L_1^2$ is,
\begin{equation}\label{statwo5a}
\left(L_1^2\right)_{min} = \frac{M_{I} R_0^2}{R_0-3M_{I}}
\end{equation}
because for this value we have $L_2^2=L_0^2$, the minimum possible value for $L_2^2$. The value (\ref{statwo5a}) coincides with the single particle stability limit, but we may also check that for this configuration, with $L_2^2=L_0^2$, we have $CN_1=0$, and the particles with angular momentum equal to $L_1$ become test particles. Similarly, the highest possible value for $L_1$ is $L_1=L_0$. Here stability requires,
\begin{equation}\label{statwo5b}
\left(L_2^2\right)_{max} = \frac{M_{II} R_0^2}{R_0-3M_{II}}
\end{equation}
which is again the single particle stability limit, but for $L_2$, with similar considerations as for $L_1$. The important general conclusion from this analysis is that the criterion for stability under shell division is {\em stronger} than that for single particle evaporation, since it leads to critical $L_i$ values that are either {\em lower} for $L_2$, or {\em higher} than $L_1$ than those required from single particle stability.

It is interesting to check also that a static shell with a single component is stable under splitting. We consider again Eqs. (\ref{statwo1a},\ref{statwo1b},\ref{statwo2a},\ref{statwo2b}), with $L_1=L_2=L_0$. From (\ref{statwo2a},\ref{statwo2b}) we find,
\begin{equation}\label{statwo7}
\sqrt{1-\frac{2M_{int}}{R_0}} = \frac{N_1}{N_1+N_2} \sqrt{1-\frac{2M_{II}}{R_0}}+\frac{N_2}{N_1+N_2} \sqrt{1-\frac{2M_{I}}{R_0}}
\end{equation}
with $N_1$, and $N_2$ restricted by,
\begin{equation}\label{statwo8}
C (N_1+N_2)=\frac{2 R_0^2}{\sqrt{R_0^2+L_0^2}}\left(\sqrt{1-\frac{2M_{I}}{R_0}}-\sqrt{1-\frac{2M_{II}}{R_0}}\right)
\end{equation}
but otherwise the ratio $N_2/N_1$ is arbitrary. Replacing (\ref{statwo7}), and $L_2=L_0$ in, e.g., (\ref{statwo1b})
\begin{equation}\label{statwo9}
 \frac{d^2 R_2}{d \tau_2^2}= \frac{ N_1 (\sqrt{R_0-2 M_{II}}-\sqrt{R_0-2 M_{I}})}{2 (N_1+N_2)R_0 \sqrt{R_0-2 M_{I}}} < 0
\end{equation}
so that the shell is stable for all ratios $N_1/N_2$.

\section{ Stability under shell division of non static two component shells.}

As indicated in the previous Section, the stability analysis in the general non static case is, even for a two component shell, considerably more  complicated in detail than in the static case. The analysis may be carried out along lines similar to those used for the stability under single particle evaporation. We first notice that for a two component shell the equations of motion are,
\begin{equation}\label{nons01}
\frac{d^2 R}{d\tau^2} = \frac{1}{2 R}\left[1 +\dot{R}^2   -\left( 4 \alpha+1\right)
\sqrt{1-\frac{2 M_{I}}{R} +\dot{R}^2}\sqrt{1-\frac{2 M_{II}}{R} +\dot{R}^2}\right]
\end{equation}
where,
\begin{equation}\label{nons02}
\alpha =\frac{1}{2(N_1\sqrt{R^2+L_1^2}+N_2\sqrt{R^2+L_2^2})}\left[\frac{N_1 L_1^2}{\sqrt{R^2+L_1^2}}+\frac{N_2 L_2^2}{\sqrt{R^2+L_2^2}}\right]
\end{equation}
and,
\begin{equation}\label{nons03}
 \sqrt{1-\frac{2 M_{I}}{R} +\dot{R}^2} - \sqrt{1-\frac{2 M_{II}}{R} +\dot{R}^2}
=
\frac{ C N_1\sqrt{R^2+L_1^2}}{2 R^2}+ \frac{ C N_2\sqrt{R^2+L_2^2}}{2 R^2}
\end{equation}
where $R=R(\tau)$, $\dot{R}=dR/d\tau$, and $\tau$ is the proper time for an observer that moves radially with the shell. Notice that (\ref{nons03}) is actually a first integral of (\ref{nons01}). A particular solution of the equations of motion is determined by fixing, for some particular value of $\tau$, say $\tau=\tau_0$,   appropriate values of $M_{I}$, $M_{II}$, ($M_{II} > M_{I}$), $L_1$, $L_2$, ($L_2 > L_1$), $C N_1 \geq 0$, and $C N_2 \geq 0$, and a corresponding radius $R(\tau_0)=R_0$, such that (\ref{nons03}) may be solved for a real value of $dR/d\tau$. We consider now, as in the previous Section, for the  same value of $\tau=\tau_0$, an infinitesimal displacement of the components, so that we have two shells, one made out of the particles with $L_1$, and the other with $L_2$, separated by an intermediate empty region with mass $M_{int}$. We assume that $M_{I}$, $M_{II}$, $L_1$,  $L_2$, $C N_1$, are $C N_2$, are not modified by this displacement, and that the shells have initially vanishing relative velocity. If we look now at the $L_2$ component, its equations of motion should be,
\begin{equation}\label{nons01a}
\frac{d^2 R_2}{d\tau_2^2} = \frac{1}{2 R_2}\left[1 +\dot{R_2}^2   -\left( 4 \alpha_2+1\right)
\sqrt{1-\frac{2 M_{int}}{R_2} +\dot{R_2}^2}\sqrt{1-\frac{2 M_{II}}{R_2} +\dot{R_2}^2}\right]
\end{equation}
where,
\begin{equation}\label{nons02a}
\alpha_2 =\frac{ L_2^2}{2(R_2^2+L_2^2)}
\end{equation}
and,
\begin{equation}\label{nons03a}
 \sqrt{1-\frac{2 M_{II}}{R_2} +\dot{R_2}^2} - \sqrt{1-\frac{2 M_{int}}{R_2} +\dot{R_2}^2}
 + \frac{ C N_2\sqrt{R_2^2+L_2^2}}{2 R_2^2} = 0
\end{equation}
where $R_2=R_2(\tau_2)$, and $\tau_2$ is now the proper time for the $L_2$ shell. We assume that for $\tau=\tau_0$ we also have $\tau_2=\tau_0$, so that   $R(\tau=\tau_0)=R_2(\tau_2=\tau_0)$. From the above assumptions and metric continuity, we also have $d R_2(\tau_2)/d\tau_2= d R(\tau)/d\tau$. Then, we may use (\ref{nons03a}) to compute $M_{int}$, and then obtain $d^2 R_2(\tau_2)/d\tau_2^2$ from (\ref{nons01a}). We might be tempted to conclude that the condition for stability of the shell should then be $d^2 R_2(\tau_2)/d\tau_2^2 < d^2 R(\tau)/d\tau^2$, but this could in principle be incorrect, because of the different meanings of the proper times $\tau$, and $\tau_2$. It turns out, however, that this is the correct criterion. We may prove this as follows. From the point of view of an observer in the region ${\cal{M}_{II}}$, the motion of the shell is described by a function $R=R(t_{II})$, and we have the following relations between the derivatives with respect to $t_{II}$ and to proper time $\tau$ (or $\tau_2$) of the shell,
\begin{equation}\label{nons04}
\left(\frac{d \tau}{d t_{II}}\right)^2 = \frac{\left(R(\tau)-2 M_{II} \right)^2}{R(\tau)^2-2 R(\tau) M_{II} +R(\tau)^2 \left(\frac{dR(\tau)}{d\tau}\right)^2}
\end{equation}
from which we may also compute $d^2 \tau/d t_{II}^2$, and $d^2 \tau_2/d t_{II}^2$. From these expressions we may finally derive,
\begin{eqnarray}
\label{nons05}
\frac{d^2 R(t_{II})}{d t_{II}^2} & = & \frac{R}{R-2 M_{II}}\left(\frac{d \tau}{dt_{II}}\right)^4 \frac{d^2 R(\tau)}{d \tau^2} \nonumber \\
   & &  + \frac{M_{II}}{R^3(R-2 M_{II})} \left[R-2 M_{II} +R \left(\frac{d \tau}{dt_{II}}\right)^2 \right]\left[2 R-4 M_{II} + R \left(\frac{d \tau}{dt_{II}}\right)^2 \right]
\end{eqnarray}
and a similar expression with $R$ replaced by $R_2$ and $\tau$ by $\tau_2$. But, from the above assumptions, for $\tau=\tau_2=\tau_0$ we have $d\tau/dt_{II}=d \tau_2/dt_{II}$, and then the shell with the larger proper radial acceleration will also be seen to have the larger radial acceleration with respect to $t_{II}$. Moreover, the critical condition, i.e., equality of the second derivatives with respect to $t_{II}$, implies also equality of the second derivatives of $R$, and $R_2$, with respect respectively to $\tau$ and $\tau_2$. If we assume that at a certain point, we have $R(\tau)=R(\tau_2)$, $dR/d\tau=dR_2/d\tau_2$, and $d^2R/d\tau^2=d^2R_2/d\tau_2^2$, an replace in (\ref{nons01}) to (\ref{nons03a}), we obtain the following condition,
\begin{equation}\label{nons06}
\frac{(3 L_2^2+R^2) }{(R^2+L_2^2)}\sqrt{1-\frac{2M_{II}}{R} + \left(\frac{dR}{d\tau}\right)^2} =\frac {(3 L_1^2+R^2) }{(R^2+L_1^2)}\sqrt{1-\frac{2M_{I}}{R}+ \left(\frac{dR}{d\tau}\right)^2}
\end{equation}
which is seen to be a simple generalization of the critical condition for stability for static shells given by (\ref{statwo5}). It is again remarkable that this is also the critical condition if we consider stability for the shell with $L_1$. Since the values of $R$ and $dR/d\tau$ change as a function of $\tau$, a shell that is initially stable may become unstable as it expands or contracts, and spontaneously split into two separate components, one with angular momentum $L_1$ and the other with $L_2$.

However, for an expanding shell, as $R$ increases, assuming given value of $L_1$, $M_{I}$, and $M_{II}$,  (\ref{nons06}) is satisfied for increasing values of $L_2$. This can be seen solving  (\ref{nons06}) for $L_2^2$, and assuming that $R$ is much larger than the other quantities in  (\ref{nons06}), (we recall that $\dot{R}=dR/d\tau$ is always bounded, even for $R \to \infty$). The result is,
\begin{equation}\label{largeR1}
L_2^2 = \frac{M_{II}-M_{I}}{2 (1+\dot{R}^2)} R +L_1^2+\frac{M_{II}(M_{II}-M_{I})}{(1+\dot{R}^2)^2}+ {\cal{O}}(R^{-1})
\end{equation}
and, therefore, for sufficiently large $R$, any $L_2$ will satisfy the stability condition. Since instability results for either $L_2$ larger (or $L_1$ smaller) than the critical value, we conclude that {\em any two component shell necessarily becomes stable if it expands to sufficiently large values of $R$}.

The derivations for the two component shells strongly suggest that these results hold also for {\em any} multi component shell. If this is correct, a thin shell with a continuous distribution of $L$ values might, upon evolution, turn into a thick shell with a continuous plus a singular distribution of matter, an interesting result in view of some theorems \cite{Rendall} on the emergence of singularities in Vlasov - Einstein systems. We shall consider this point in more detail in a separate paper.

Going back to the stability problem, in principle, given $M_{I}$, $M_{II}$, $L_1$, $L_2$, $CN_1$, and $CN_2$, we may solve (\ref{nons03}) for $(dR/d\tau)^2$, and replace in (\ref{nons06}) to obtain an equation for $R$ in terms of these constants, whose solutions contain the values of $R$ where the shell becomes unstable. Unfortunately, the explicit expressions are too complicated to allow for a simple interpretation of their meaning. Instead, we consider several explicit examples where we integrate (\ref{nons01}), and use the resulting $R(\tau)$, and $dR/\tau$ to compute $M_{int}$, and then the right hand side of (\ref{nons05}), also as  a function of $\tau$. The shell will become unstable if this is larger than $d^2 R/d\tau^2$ in (\ref{nons01}). Similarly with regards to the shell with $L_1$, except that here the shell becomes unstable when the corresponding acceleration is smaller than  $d^2 R/d\tau^2$, but we must emphasize that, as shown above, the instability condition is always simultaneously satisfied for both component shells.

As a first example we consider a solution of the shell equations that corresponds to a bounded periodic motion. For this example we chose $M_{I}=0.45$, $M_{II}=1$, $C=1$, $L_1=2.8$,$L_2=3.7$, $N_1=0.3119$, and $N_2=0.8044$. Figure 1 shows $R(\tau)$ as a function of $\tau$ for this choice of parameters. We have two turning points, one for $R=8.$, and the other for $R=40.0...$. The period is $\Delta \tau \simeq 940$. We have chosen $\tau=0$ at the lower turning point $R=8$.

\begin{figure}
\centerline{\includegraphics[height=8cm,angle=0]{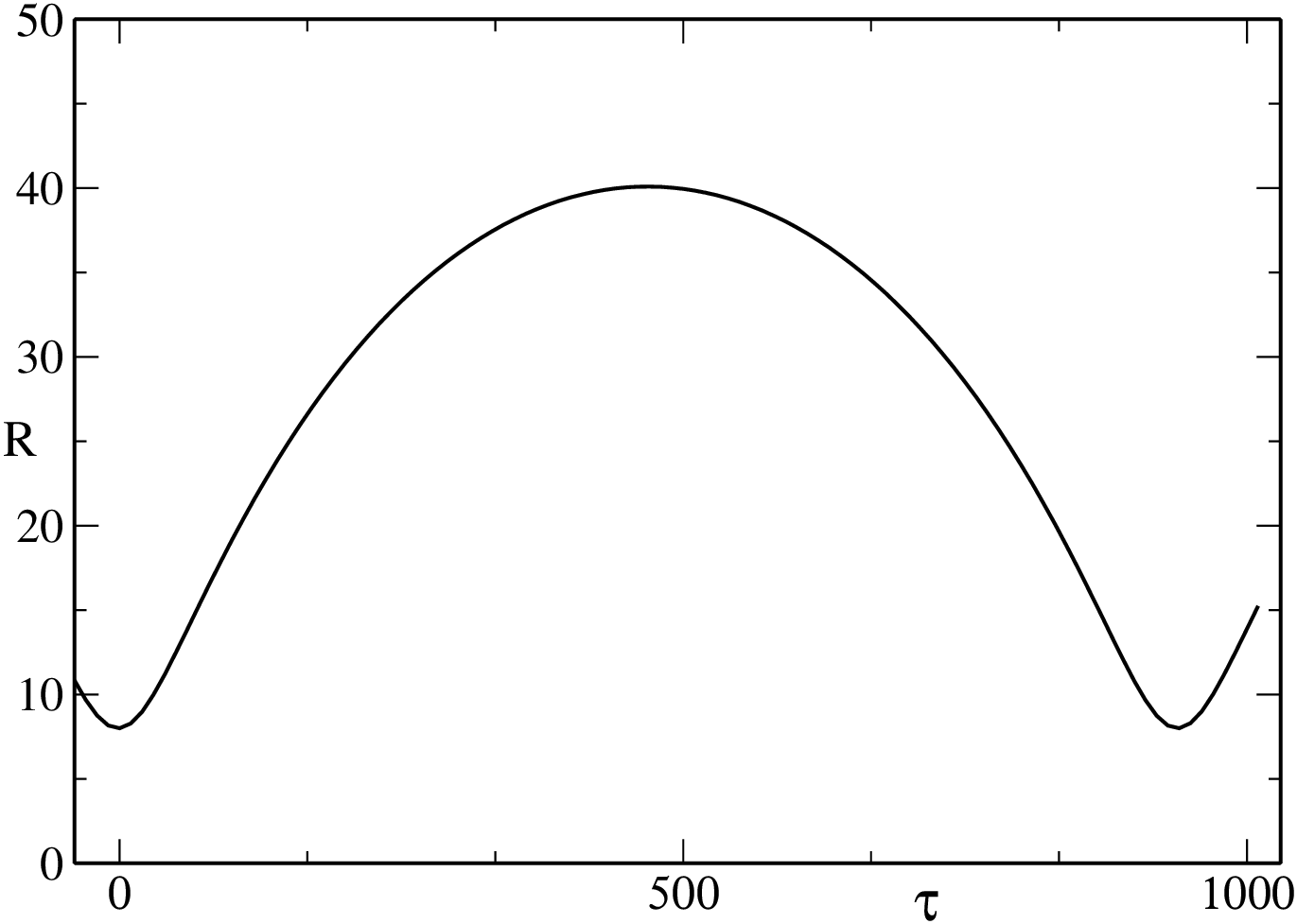}}
\caption{$R(\tau)$ as a function of $\tau$ for $M_{I}=0.45$, $M_{II}=1$, $C=1$, $L_1=2.8$,$L_2=3.7$, $N_1=0.3119$, and $N_2=0.8044$. The motion is periodic and bounded.}
\end{figure}

To analyze the stability under shell splitting, we computed $d^2 R_2/d\tau^2_2$, and $d^2 R_1/d\tau^2_1$, as functions of $\tau$ as indicated above, and compared their values with those of $d^2 R/d\tau^2$. The relevant results are indicated in Figure 2, where curve $a$ corresponds to $d^2 R/d\tau^2$, curve $b$  to $d^2 R_2/d\tau^2_2$, and $c$  to $d^2 R_1/d\tau^2_1$. The curve is restricted to $50 \leq \tau \leq 140$ to show the point where the shell turns from a stable to an unstable motion, at the critical value $\tau \simeq 86$, corresponding to $R \simeq 17.$. The motion is unstable to the left and stable to the right of this point. Given the fact that the motion of the shell as a whole is periodic and invariant under $\tau \to -\tau$, we conclude that the motion is stable under shell splitting for $R > \sim 17$ and unstable for $R < \sim 17$. We must remark. however, that the if the shell starts its motion in the stable region, and a splitting does occur at the critical point, the ensuing motion will no longer be periodic, because the resulting separate shells evolve with different proper times, and, in general, they will have a non vanishing relative velocity when they cross each other again. This kind of motions has been investigated by Barkov, et.al., \cite{Barkov} who find a very complex chaotic like pattern for the motions of the separate shells. Thus, in this example, a highly ordered initial state, corresponding to a single shell, may turn into a very complex, non periodic motion of the separate shells. The analysis of the motion of the shells after separation is, however, outside the scope of the present analysis.

\begin{figure}
\centerline{\includegraphics[height=8cm,angle=0]{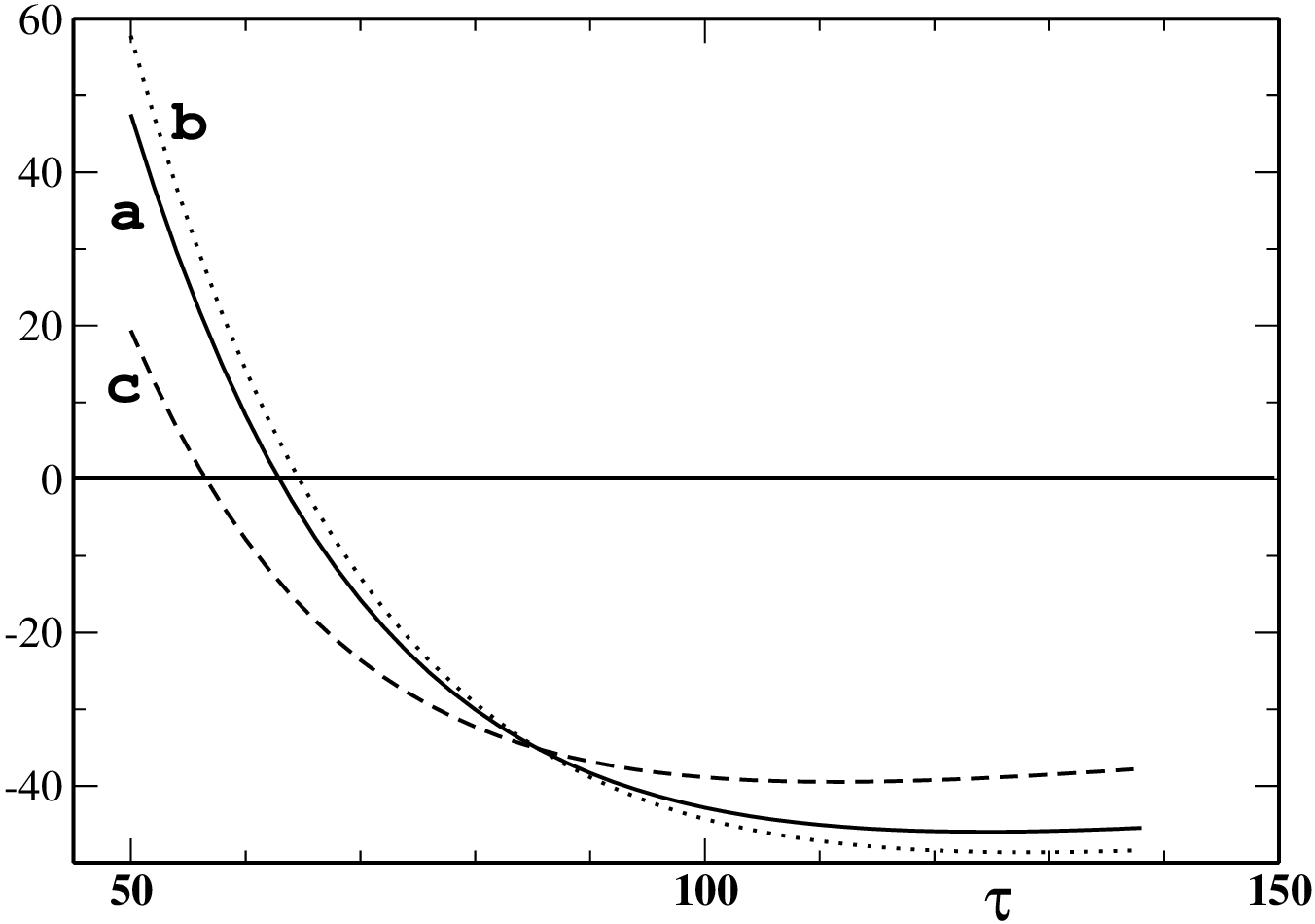}}
\caption{Curve $a$ (solid line) corresponds to the acceleration $d^2R/d\tau^2$ of the shell, curve $b$ (dotted line), to $d^2R_2/d\tau_2^2$, computed as the right hand side of (\ref{nons01a}), and curve $c$ (dashed line), to $d^2R_1/d\tau_1^2$, similarly computed. Only the region where the shell becomes critical is depicted. The critical value of $\tau$ corresponds to the crossing of the curves, where the three accelerations are equal. To the right of this point we have $d^2R_2/d\tau_2^2 < d^2R/d\tau^2$, and $d^2R_1/d\tau_1^2 > d^2R/d\tau^2$, and the shell is stable. The opposite situation holds for $\tau$ less than the critical value. A vertical unit in this graph corresponds to $10^{-5}$.}
\end{figure}

In general, it is also possible to find choices of parameters such that, for bounded periodic motions, the shell is unstable at all points, or, on the contrary, it is stable at all points.

As a second example we take $M_{I}=0.6295$, $M_{II}=1$, $C=1$, $L_1=10$,$L_2=11$, $N_1=0.2$, and $N_2=0.3$. Figure 3 shows $R(\tau)$ as a function of $\tau$ for this choice of parameters. We have a single turning point for $R=8.$, and we have chosen $\tau=0$ at this point. Here we have two critical points, one at $\tau \simeq 6.03...$ corresponding to $R(\tau)\simeq 9.10...$, (see Fig. 4) and the other for $\tau \simeq 232.5...$, corresponding to $R(\tau)\simeq 246.7...$ (see Fig. 5). The motion is unstable for $R$ in this range, but stable near the turning point, and for $R$ larger than the larger critical value, in accordance with our previous discussion on the behaviour for large $R$.

\begin{figure}
\centerline{\includegraphics[height=8cm,angle=0]{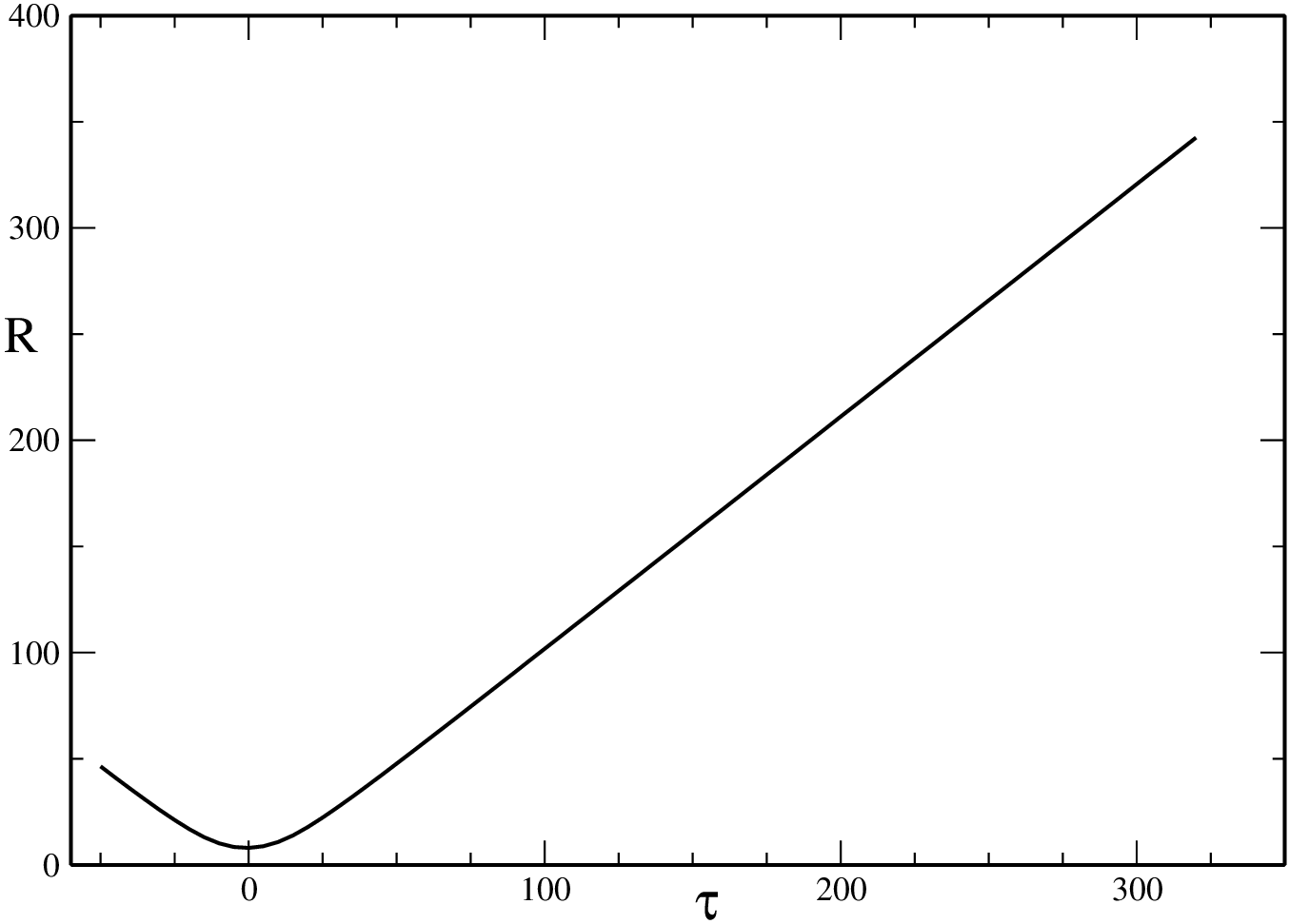}}
\caption{$R(\tau)$ as a function of $\tau$ for $M_{I}=0.6295$, $M_{II}=1$, $C=1$, $L_1=10$,$L_2=11$, $N_1=0.2$, and $N_2=0.3$. The motion has a lower turning point at $R=8$, but is unbounded.}
\end{figure}

\begin{figure}
\centerline{\includegraphics[height=8cm,angle=0]{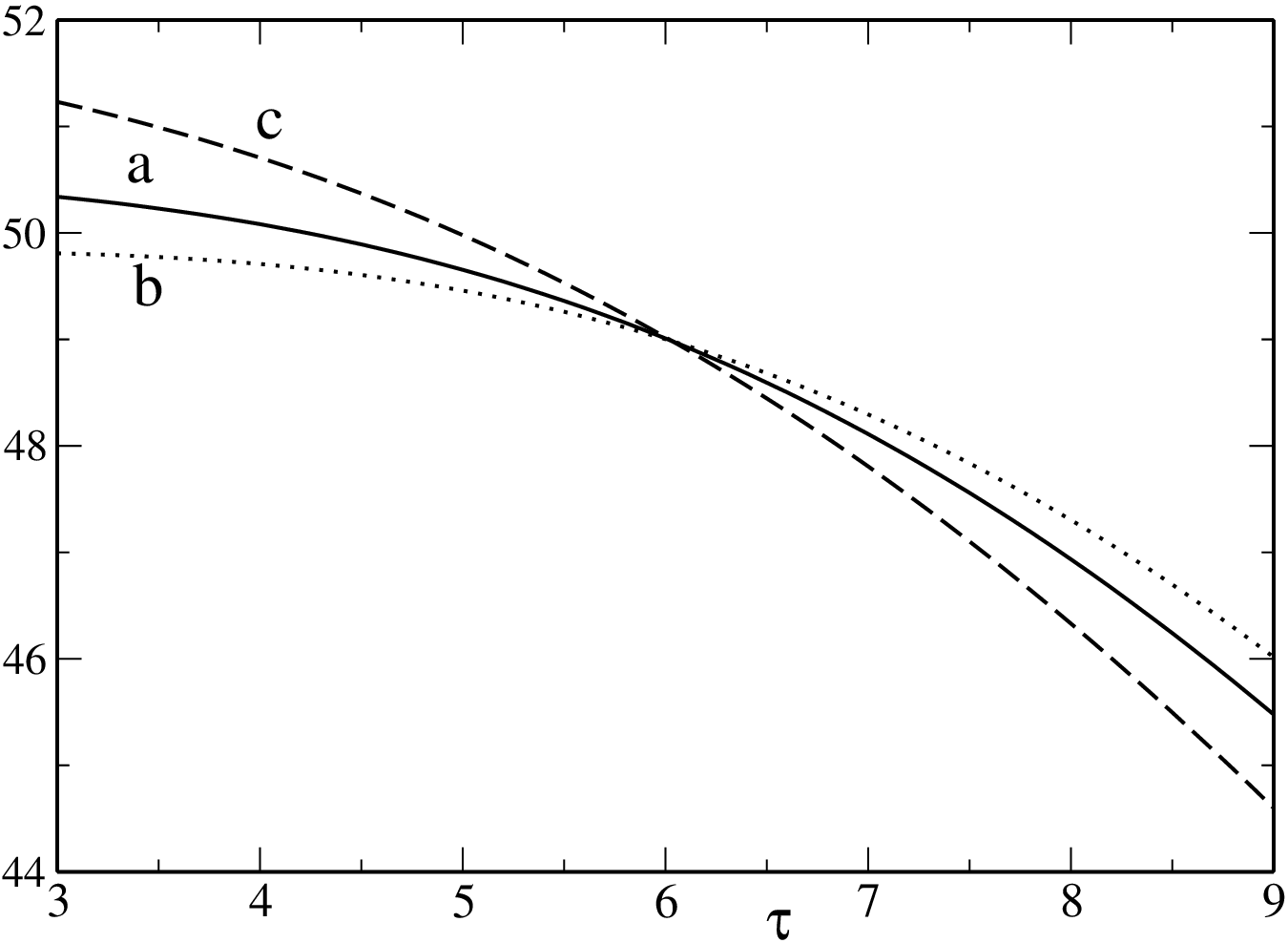}}
\caption{Curve $a$ (solid line) corresponds to the acceleration $d^2R/d\tau^2$ of the shell, curve $b$ (dotted line), to $d^2R_2/d\tau_2^2$, computed as the right hand side of (\ref{nons01a}), and curve $c$ (dashed line), to $d^2R_1/d\tau_1^2$, similarly computed, and corresponding to the parameters in Fig. 3. Only the region near the critical point at $\tau \simeq 6.03...$ is depicted. The critical value of $\tau$ corresponds to the crossing of the curves, where the three accelerations are equal. To the left of this point we have $d^2R_2/d\tau_2^2 < d^2R/d\tau^2$, and $d^2R_1/d\tau_1^2 > d^2R/d\tau^2$, and the shell is stable. The opposite situation holds for $\tau$ to the right of this critical value (but less than the higher critical value displayed in Fig. 5). A vertical unit in this graph corresponds to $10^{-3}$.}
\end{figure}

\begin{figure}
\centerline{\includegraphics[height=8cm,angle=0]{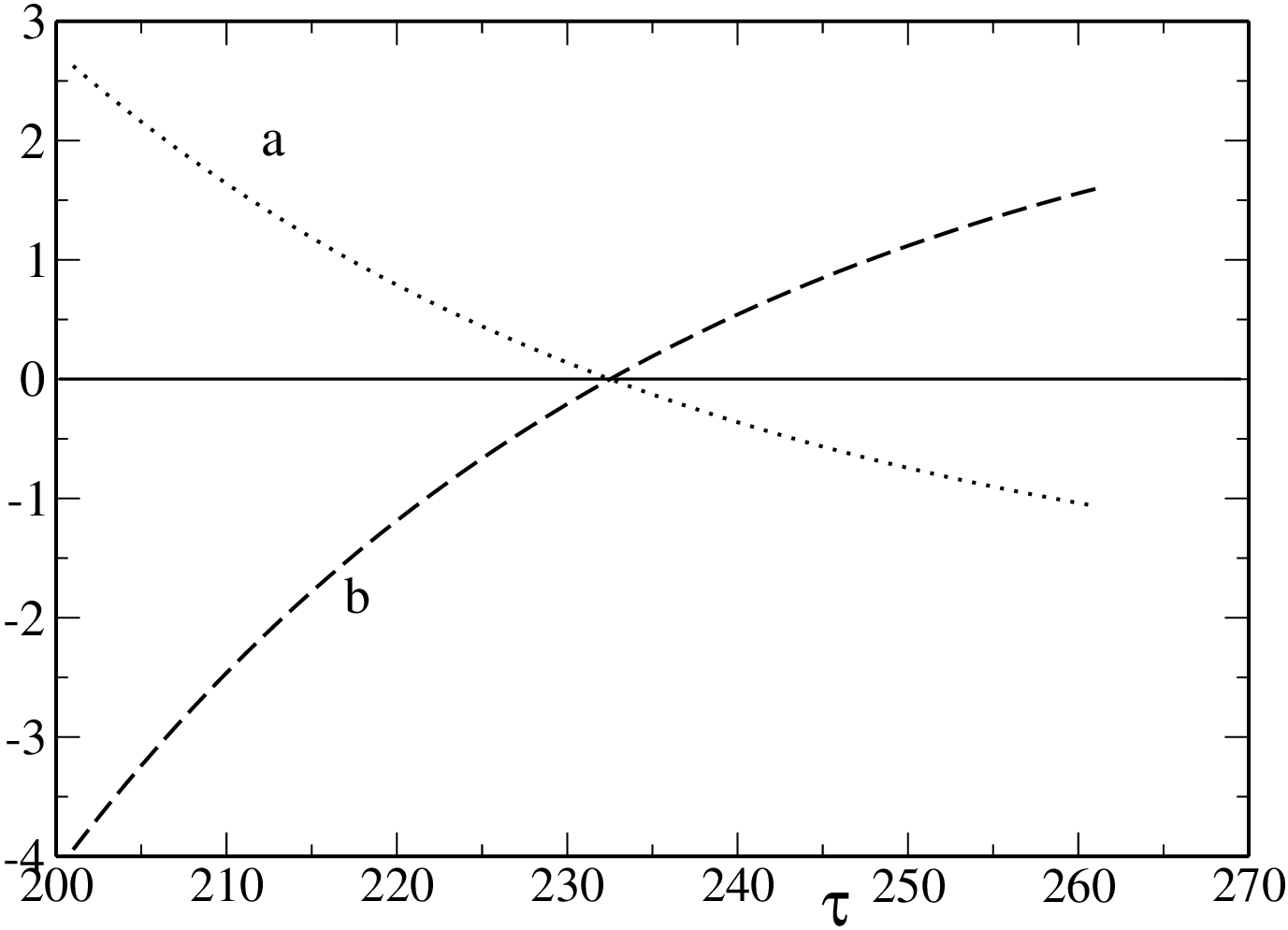}}
\caption{Curve $a$ (dotted line) corresponds to  $d^2R_2/d\tau_2^2-d^2R/d\tau^2$, and curve $b$ (dashed line), to $d^2R_1/d\tau_2^1-d^2R/d\tau^2$, computed as in Fig. 4, and corresponding to the parameters in Fig. 3. Only the region near the critical point at $\tau \simeq 232.5...$ is depicted. The critical value of $\tau$ corresponds to the crossing of the curves, where the three accelerations are equal. To the right of this point we have $d^2R_2/d\tau_2^2 < d^2R/d\tau^2$, and $d^2R_1/d\tau_1^2 > d^2R/d\tau^2$, and the shell is stable. The opposite situation holds for $\tau$ less than this critical value (but larger than the lower critical value displayed in Fig. 4). A vertical unit in this graph corresponds to $10^{-7}$.}
\end{figure}

\section{Comments and conclusions.}

One question that arises in trying to interpret the instabilities we have found in this paper is why do we have this ambiguity at the critical points, and unstable regions, where, both a single shell and a split pair are compatible with the equations of motion for the particles, that are explicitly assumed to be collisionless, and interact only gravitationally. A possible answer is that the equations of motion for the shell contain implicitly a ``constraint'' that forces the particles to remain on the shell. The effect of the constraint is irrelevant in the stable region, where, at least in the Newtonian limit, one can see a ``wedge'' shaped potential, keeping the particles at the bottom of the wedge, but is crucial in the unstable region, where the wedge can no longer restrain the particles. This ``constraint'', would have to correspond to a massless but both stiff and elastic structure, and , therefore, we consider the splitting interpretation as the more physical one in our case. This would clearly have to be taken in consideration in any application to physically meaningful systems.

\appendix*

\section{The thick to thin shell limit for static Einstein shells}

We consider a static thick Einstein shell, that is, a spherically symmetric space time where the matter contents is made out of equal mass, non interacting particles moving along circular geodesics, and confined to a spherical shell of non vanishing thickness. The metric for this space time may be written in the form,
\begin{equation}\label{app01}
ds^2= -e^{\nu(r)} dt^2 + \left(1 -\frac{2 m(r)}{r} \right)^{-1} dr^2 + r^2 \left(d \theta^2 +\sin^2\theta d\phi^2\right)
\end{equation}
We assume a central source of mass $M_{I}$, and that the shell is restricted to $R_i \leq r \leq R_o$, with $2M_{I} < R_i$. Then, for $ 2M_{I} < r < R_i$ we have,
\begin{equation}\label{app02a}
m(r)=M_{I} \;\;\; ; \;\;\; e^{\nu(r)}= 1 -\frac{2 M_{I}}{r}
\end{equation}
Because of the positivity of the energy density, (see, Eq. (\ref{app07})), the function $m(r)$ is increasing as a function of $r$ in the interval $R_i\leq r \leq R_o$, (inside the shell), attaining its maximum value $m(r) = M_{II}$, at the outer boundary $r=R_o$ of the shell. Then, for $r \geq R_o$ we have,
\begin{equation}\label{app02b}
m(r)=M_{II} \;\;\; ; \;\;\; e^{\nu(r)}= C \left(1 -\frac{2 M_{II}}{r}\right)
\end{equation}
where $C$ is a constant.

The stress-energy-momentum tensor is given by,
\begin{equation}\label{app03}
T_{ab}(x^{\sigma}) = \mu^{-1}\, n(r) \langle p_{a} p_{b}\rangle_{avg}
\end{equation}
where $\mu$ is the mass of the particles, $n(r)$ is proportional to the proper particle number density, and the particle 4-momenta $p_{a}$
are averaged over all space directions, compatible with the condition of circular geodesic
motion. Since $p^r=0$ for all particles, this implies that all components of the form $T_{r\, a}$ vanish. In particular, imposing $T_r{}^r=0$ on Einstein's equations we find,
\begin{equation}\label{app05}
\frac{d\nu}{dr} = \frac{2 m(r)}{r(r-2 m(r))}
\end{equation}

Moreover, the condition of circular orbits implies that for all particles $p^{a}$ satisfies,
\begin{equation}\label{app06}
p_{a} p^{a} = -\mu^2 = -e^{\nu(r)} \left(p^t\right)^2 + \frac{\mu^2 L^2}{ r^2}
\end{equation}
where $L$ is the particle's angular momentum per unit mass. A simple computation then shows that the only non vanishing
components of $T_{ab}$ are,
\begin{eqnarray}
\label{app04a}
T_t{}^t &=& - \mu n(r)\left[1 +\frac{L^2}{r^2}\right] = -  \rho \\
\label{app04b}
T_{\theta}{}^{\theta} &=& T_{\phi}{}^{\phi} = \mu \frac{n(r) L^2}{2 r^2}
\end{eqnarray}
where $\rho$ is the energy density. From Einstein's equations we have,
\begin{equation}\label{app07}
8 \pi \rho = \frac{2}{r^2} \frac{dm}{dr}
\end{equation}

We shall be interested in the thin shell limit, where $R_i \to R_o \to R_0$, where $R_0$ is some radius. Then, unless the shell is empty, $\rho$ approaches a Dirac's $\delta$ form, and so does $n(r)$, which is therefore singular in this limit. Nevertheless, as we shall show, if instead of considering the number of particles at a given $r$, we look at the distribution of values of $L^2$, we find that as the shell becomes thinner and thinner, this distribution approaches a unique smooth form, that depends only on $M_{I}, M_{II}$, and, and the radius $R_0$ of the limiting thin shell. This can be seen as follows.

The condition that the world lines of the particles be geodesics of the metric (\ref{app01}) imposes that,
\begin{equation}\label{app08}
L^2 = \frac{ r^2 m(r)}{r-3 m(r)}
\end{equation}
which makes explicit the condition that no part of the shell may have $r\leq 3 m(r)$. In particular, we must have $R_o \geq 3 M_{II}$, so that thick shells cannot be made more compact than thin shells. From (\ref{app08}) we get,
\begin{equation}\label{app11}
\frac{dL^2}{dr} = \frac{ r^3}{(r-3m)^2}\frac{dm}{dr} +\frac{ r m (r-6m)}{(r-3m)^2}
\end{equation}
and therefore $dL^2/dr > 0$ for $r > 6m(r)$, or $dm/dr$ sufficiently large. We will be interested in the case where $R_i \to R_o$, where $dm/dr$ becomes arbitrarily large, since $m(r)$ grows from $M_{I}$ to $M_{II}$ in an arbitrarily small interval of $r$. In this case, we may assume $dL^2/dr > 0$, and therefore, we have that $L^2$ grows monotonically in the interval,
\begin{equation}\label{app12}
\frac{ R_i^2 M_{I}}{R_i-3 M_{I}} \leq L^2 \leq \frac{ R_o^2 M_{II}}{R_o-3 M_{II}}
\end{equation}

Let now $\tilde{n}(r) dr$ be the total number of particles between $r$ and $r+dr$. Then we have,
\begin{equation}\label{app10}
4 \pi \rho r^2 \left( 1 - \frac{2 m(r)}{r}\right)^{-1/2}  = \mu \tilde{n}(r) \sqrt{1 +\frac{L^2}{r^2}}
\end{equation}

Since we assume $dL^2/dr > 0$, we may introduce,
\begin{equation}\label{app13}
\hat{n}(L^2) =  \tilde{n}(r) \frac{dr}{dL^2}
\end{equation}
where $\hat{n}(L^2) d L^2$ is the total number of particles in the shell with (squared) angular momentum in $(L^2, L^2 +dL^2)$. We may solve (\ref{app08}) for $m(r)$,
\begin{equation}\label{app09}
m(r) = \frac{L^2 r}{3 L^2+ r^2}
\end{equation}
and derive with respect to $L^2$ to get,
\begin{equation}\label{app14}
\frac{dm}{dL^2} = \frac{ r^3}{(3L^2+ r^2)^2} +{\cal{O}}\left(\frac{dr}{dL^2}\right)
\end{equation}
and we remark that ${dr}/{dL^2} \to 0$ as $R_i \to R_o$. We may now solve (\ref{app10}) for $\tilde{n}$, and using the previous results, we find,
\begin{equation}\label{app15}
\hat{n}(L^2)= \frac{ R_0^4}{\mu (3L^2+ R_0^2)^{3/2}(L^2+ R_0^2)}
\end{equation}
where we have taken the thin shell limit $R_i \to R_o \to R_0$, and $L^2$ is restricted to the range,
\begin{equation}\label{app12a}
\frac{ R_0^2 M_{I}}{R_0-3 M_{I}} \leq L^2 \leq \frac{ R_0^2 M_{II}}{R_0-3 M_{II}}
\end{equation}

Thus we see that in the thin shell limit, an Einstein shell contains particles with a unique continuous distribution of angular momentum given by $\hat{n}(L^2)$, with $L^2$ varying continuously in the range (\ref{app12a}).


\section*{Acknowledgments}

This work was supported in part by grants from CONICET (Argentina)
and Universidad Nacional de C\'ordoba.  RJG and MAR are supported by
CONICET.


\begin{thebibliography}{99}
\bibitem{Andreas} See, for instance, H.Andr\'easson, The Einstein–Vlasov system/Kinetic theory \textit{Living Rev. Rel.} \textbf{8}, 2 (2005) http://www.livingreviews.org/lrr-2005-2, and references therein.
\bibitem{einstein} A.Einstein, \textit{Ann. Math.}
\textbf{40}, 922(1939)
\bibitem{Hamity} V. H. Hamity, ``Relativistic spherically symmetric thin-shelled ensenmbles of collisionless particles in the presence of acentral boby'', in \textit{Proceedings of SILARG III}, Universidad Nacional Aut\'onoma de M\'exico, M\'exico, 1982
\bibitem{Berezin} V Berezin and M Okhrimenko, \textit{Class. Quantum Grav.} \textbf{18}, 2195 (2001)
\bibitem{Evans} A.B.Evans \textit{Gen. Rel. Grav.} \textbf{8},155 (1977)
\bibitem{Israel} W.Israel \textit{Nuov. Cim.} B\textbf{44}, 1 (1966)
\bibitem{Barkov} M. V. Barkov, V. A. Belinski, and G. S. Bisnovatyi-Kogan, \textit{Journal of Experimental and Theoretical Physics}, Vol. \textbf{95}, 371, (2002).
\bibitem{Lobo} See, for instance, F S N Lobo and P Crawford, \textit{Class. Quantum Grav.} \textbf{22}, 4869 (2005), and references therein.
\bibitem{AndreasBuch} H. Andr\'easson, \textit{On static shells and the Buchdahl inequality for the spherically symmetric Einstein-Vlasov system}, [arXiv:gr-qc/0605151v1]
\bibitem{ComKatz} G.L.Comer, and J.Katz \textit{Class. Quant. Grav.} \textbf{10}, 1751 (1993)
\bibitem{Rendall} See, for intance, A. D. Rendall, ``The Einstein-Vlasov system.'', [arXiv:gr-qc/0208082]
\end{thebibliography}
\end{document}